%% file: panic.tex
\documentclass[english,a4paper,prb,reprint,amsmath,amssymb,superscriptaddress,nofootinbib]{revtex4-1}

\usepackage{hyperref}
\usepackage{graphicx}
\usepackage{epstopdf}
\usepackage[caption=false]{subfig}
\usepackage{booktabs}
\usepackage{xcolor,soul}

\usepackage{siunitx}
\DeclareSIUnit{\litre}{l}
\newcommand{\LineSIrange}[3]{\SIrange[range-phrase=--]{#1}{#2}{#3}}

\usepackage{bm}

\usepackage{upgreek}

\newcommand{\um}{\micro\metre}

\newcommand{\eg}{e.g.\ }
\newcommand{\ie}{i.e.\ }
\newcommand{\cf}{cf.\ }

\newcommand{\Refcite}[1]{Ref.~\onlinecite{#1}}
\newcommand{\Refscite}[1]{Refs.~\onlinecite{#1}}
\newcommand{\Figref}[1]{Fig.~\ref{#1}}

\graphicspath{{Figures/},{Figures/SEM/}}

\definecolor{BLUE}{RGB}{0,0,255}
\newcommand{\joshspeaks}[1]{\textcolor{blue}{#1}}
\renewcommand{\joshspeaks}[1]{#1}

\usepackage{footnote}

\begin{document}

\title{How effective are face coverings in reducing transmission of COVID-19?}
\date{\today}

\author{Joshua F. Robinson}
\email{joshua.robinson@uni-mainz.de}
\affiliation{H.\ H.\ Wills Physics Laboratory, University of Bristol, Bristol BS8 1TL, United Kingdom}
\affiliation{Institut f\"ur Physik, Johannes Gutenberg-Universit\"at Mainz, Staudingerweg 7-9, 55128 Mainz, Germany}
\author{Ioatzin Rios de Anda}
\affiliation{H.\ H.\ Wills Physics Laboratory, University of Bristol, Bristol BS8 1TL, United Kingdom}
\affiliation{School of Mathematics, University Walk, University of Bristol, BS8 1TW, United Kingdom}
\author{Fergus J. Moore}
\affiliation{H.\ H.\ Wills Physics Laboratory, University of Bristol, Bristol BS8 1TL, United Kingdom}
\affiliation{School of Mathematics, University Walk, University of Bristol, BS8 1TW, United Kingdom}
\author{Florence K. A. Gregson}
\affiliation{School of Chemistry, Cantock's Close, University of Bristol, Bristol, BS8 1TS, United Kingdom}
\author{Jonathan P. Reid}
\affiliation{School of Chemistry, Cantock's Close, University of Bristol, Bristol, BS8 1TS, United Kingdom}
\author{Lewis Husain}
\affiliation{Institute of Development Studies, Library Road, Brighton, BN1 9RE, United Kingdom}
\author{Richard P. Sear}
\affiliation{Department of Physics, University of Surrey, Guildford, GU2 7XH, United Kingdom}
\author{C. Patrick Royall}
\affiliation{Gulliver UMR CNRS 7083, ESPCI Paris, Universit\'{e} PSL, 75005 Paris, France}
\affiliation{H.\ H.\ Wills Physics Laboratory, University of Bristol, Bristol BS8 1TL, United Kingdom}
\affiliation{School of Chemistry, Cantock's Close, University of Bristol, Bristol, BS8 1TS, United Kingdom}
\affiliation{Centre for Nanoscience and Quantum Information, University of Bristol, Bristol BS8 1FD, United Kingdom}

\begin{abstract}
  In the COVID--19 pandemic, billions are wearing face masks, in both health care settings and in public.
  Which type of mask we should wear in what situation, is therefore important.
  There are three basic types: cotton, surgical, and respirators (\eg FFP2, N95 and similar).
  All are essentially air filters worn on the face.
  Air filtration is relatively well understood, however, we have almost no direct evidence on the relative role played by aerosol \joshspeaks{particles} of differing sizes in disease transmission.
  But if the virus concentration is assumed independent of aerosol \joshspeaks{particle} size, then most virus will be in \joshspeaks{particles} $\gtrsim \SI{1}{\um}$. \joshspeaks{We develop a} model \joshspeaks{that} predicts surgical masks are effective at reducing the risk of airborne transmission because the filtering material most surgical masks use is highly effective at filtering particles with diameters $\gtrsim \SI{1}{\um}$. However, surgical masks are significantly less effective than masks of FFP2, N95 and similar standards, mostly due to the poor fit of surgical masks. Earlier work found that $\sim 30\%$ of the air bypasses a surgical mask and is not filtered. This highlights the fact that standards for surgical masks do not specify how well the mask should fit, and so are not adequate for protection against COVID-19.
\end{abstract}

\maketitle

\section{Introduction}

The COVID-19 pandemic has brought critically neglected areas of infection control onto the global stage \cite{dancer2020}.
Most notably this includes the risk of airborne transmission
and the strategies required to mitigate it \cite{morawska2020}.
The airborne route involves transmission of viral material through aerosols, and is the dominant transmission route for SARS-CoV-2 \cite{prather2020}.
Respiratory aerosol \joshspeaks{particles} vary in diameter from $~\sim$\SIrange{0.1}{100}{\um} \cite{bar-on2020,pratherAirborne2020}.
Given this broad size range, we can expect their potency as disease vectors and the effectiveness of interventions to be size-dependent.

Face coverings are mandated (or strongly encouraged) around the world in healthcare settings and public spaces \cite{masks4allCountries,yougov2020}.
The current evidence suggests that they reduce airborne transmission of SARS-CoV-2 \cite{greenhalgh2020,delve2020,howard2020}.
An important question concerns what kind of mask should be worn in each situation?
Different levels of protection are required in healthcare settings than out in the wider community, and requirements will further depend on the particular circumstances.
For example, \citeauthor{jonesBristol2020} \cite{jonesBristol2020} found that healthcare workers in critical care had {\em lower} than average infection rates, suggesting that other healthcare workers may be underprotected.
There are essentially three types of face coverings: (i) fabric or cloth coverings, (ii) surgical masks, and (iii) respirators (\eg N95/KN95/FFP2 or similar).
Here we use `mask' and `face covering' interchangeably to refer to any of these.
There are also masks designed to be resistant to oil aerosols, and to be splash resistant. We do not consider these here.

Here, we explore the aerosol size-dependent factors affecting mask effectiveness in aerosol \joshspeaks{particles} $\lesssim$\SI{10}{\micro\metre} in diameter.

\subsection{Aerosol dynamics and transmission}

Much work has focused on how far aerosols are transmitted, with a particular focus on establishing guidelines for physical distancing.
Multiple studies have examined the transport of respired aerosols in still air \cite{xie2007,bourouiba2014,liuCoughJets2017,bazant2020}, and other studies have explored the effect that masks have on exhalation plumes \cite{tang2009,dbouk2020,viola2021,li2021}.
Aerosol \joshspeaks{particle} size is a crucial factor affecting their distance travelled, but the focus in such studies is typically on the striking dynamical change that occurs for coarse aerosol \joshspeaks{particles} with diameters in the range $\sim$\SIrange{10}{100}{\micro\metre}.
With these large \joshspeaks{particles}, fragmentation as they pass through masks is a particular concern \cite{bourouiba2014,scharfman2016,mittal2020,poon2020,fischer2020,sharma2021}.

However, finer aerosol \joshspeaks{particles} may be disease vectors.
Fine respiratory \joshspeaks{particles} arise from within the lower respiratory tract \cite{johnson2011}, where SARS-CoV-2 pathogenesis is known to occur \cite{jin2020,kanimozhi2021}.
Viral RNA has been found in aerosol \joshspeaks{particles} smaller than $\sim$\SI{10}{\micro\metre}, though finding viable virus titers seems to depend on \joshspeaks{particle} size \cite{santarpia2021}.
One hamster study found that the majority of virus was contained in aerosol \joshspeaks{particles} smaller than \SI{5}{\micro\metre} \cite{hawks2021}.
These fine \joshspeaks{particles} have such long persistence times \cite{vuorinen2020} that how far they travel in plumes is less important than factors such as masking and air ventilation in indoor environments, where filtration is size-dependent.

A widespread model for airborne transmission in indoor environments is the Wells-Riley equation \cite{riley1978,gammaitoni1997}, which has been adapted to assess the risk of infection with SARS-CoV-2 \cite{jimenez2020,buonanno2020,dai2020}.
This model considers the limit where air in a room is well-mixed, \ie ignoring any currents or expiratory jets, while still considering such factors as air ventilation, particle deposition rates, and the rate of release of so-called \emph{infection quanta} into the room.
An infection quantum is a theoretical airborne dose expected to infect $~\SI{63}{\percent}$ of susceptible people.
Transmission risk in these models is highly dependent on the rate \joshspeaks{$q$ at which} an infected person releases \joshspeaks{these} quanta.
Given the multitude of factors involved in airborne transmission, biological and otherwise, there is large uncertainty in $q$ for the new SARS-CoV-2 pathogen \cite{buonanno2020,dai2020}.
One novel study examined the effect that distance, ventilation and masks have on infection risk by performing CFD simulations on a realistic 3d representation of a classroom \cite{foster2021}, finding reasonable overall agreement with Wells-Riley modelling despite significant deviations.

Particle size-dependence is incorporated into the Wells-Riley models (and its derivatives) through the particle deposition rates and the $q$ factor, whereas mask effectiveness has been incorporated solely as a dilution parameter as in \eg \Refcite{dai2020}.
Masks are in effect assumed to act solely by modifying the other parameters, creating an effective ventilation rate and an effective $q$ \cite{dai2020}.
The complex interplay between mask effectiveness and biological factors such as aerosol production and viral load, and how this varies with particle size, has not yet been incorporated into these models.

\subsection{Mask standards}

\begin{figure*}
  \includegraphics[width=\linewidth]{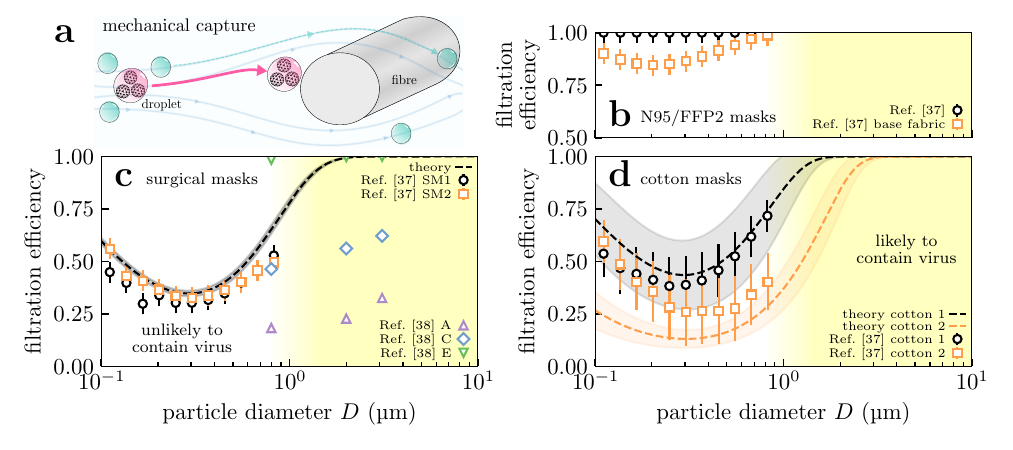}
  \caption{
    (colour online)
    Variation in mask filtration efficiency with incoming particle size.
    (a) Diagram of capture of viral particles by a fibre within a mask.
    Larger particles are more easily captured because they are less mobile; smaller aerosol \joshspeaks{particles} by contrast are transported around the fibre by the gas flow.
    Larger particles can also carry more virions, and submicron aerosol \joshspeaks{particles} are unlikely to contain even a single virion (\cf text).
    The filtration efficiency of perfectly fitting (b) N95/FFP2 masks, (c) surgical masks and (d) cloth masks formed from 4 identical plain-woven cotton layers are shown as a function of particle size.
    We show experimental measurements from \Refscite{zangmeister2020,oberg2008} (points) and predictions from our model (lines) which does not use fitting parameters and is described elsewhere \cite{robinson2021}.
    The shaded envelopes around the lines in (c-d) show the uncertainty in the model predictions, obtained by propagating uncertainties in the geometric parameters given in \Refcite{zangmeister2020}.
    We set the velocity of the gas through the mask to \SI{6.3}{\centi\metre\per\second} in our calculations for comparison with data from \Refcite{zangmeister2020}.
  }
  \label{fig:fabric-filter-efficiency}
\end{figure*}

A mask is nothing more or less than an air filter worn on the face.
Various governments have introduced minimum standards for masks. Here, we will briefly outline two European and two American standards.

For each type of face covering there are different standards. Most fabric coverings are not made to a standard. However, ASTM International recently introduced a \lq barrier face covering standard\rq ~F3502 \cite{F3502}. This specifies filtration in terms of a polydisperse aerosol of sodium chloride crystals with sizes around $\SI{100}{\nano\metre}$, but does not specify how well the mask must fit the face to avoid air leaking around the edges of the mask.

For surgical masks, standards typically only cover the filtration of the material the masks is made from. While for respirators, the standards cover both the filtration efficiency of the material \emph{and} how well it actually performs \emph{while worn}: the standards therefore specify the quality of the mask fit \cite{EuropeMaskStandard,USMaskStandard}.
As an example, we can look at the European standard EN 149 --- \lq Filtering Halfmasks to protect against particles\rq, for a filtering facepiece respirator (FFP) standard. This has three levels with increasingly stringent requirements: FFP1, 2 and 3. The FFP2 standard requires (simply speaking) \cite{EuropeMaskStandard}:
\begin{enumerate}
    \item The filtering material must filter out at least 94\% by mass of a test aerosol.
    \item The average filtration of test subjects wearing the mask (while performing standard tasks) must be at least 92\% of the mass of the test aerosol.
\end{enumerate}
The test aerosol for European standards of filtering facepiece respirator (FFP), is specified by the European standard EN 13274-7 \cite{EuropeMaskStandard}. A convenient aerosol of sodium chloride crystals is used where \lq the number median of the particle size distribution is between a diameter of
60 and \SI{100}{\nano\metre},
with a geometric standard deviation between 2.0 and 3.0\rq \cite{EuropeMaskStandard,zoller2021}. The fraction filtered is assessed by measuring the mass of sodium, \ie the mass (rather than number) fraction filtered is assessed.

Note that the USA N95 standard uses light scattering from aerosol particles to measure filtration. The aerosol specification, is also slightly different \cite{USMaskStandard}. Thus, although the USA and European methods are similar, the filtration numbers are not exactly comparable.

The European standard for surgical masks is EN 14683 --- \lq\ Medical face masks --- Requirements and test methods\rq \cite{edana}. The only filtration requirement is that the material of the mask should filter an aerosol of particles containing the bacterium {\em Staphylococcus aureus}. The bacterial filtration
efficiency (BFE) of the mask material is the fractional reduction in the number of colony forming units  (CFUs) when the aerosol is passed through the material. For a Type II mask under this standard, the BFE must \joshspeaks{achieve a} CFU reduction of at least 98\%. The aerosol is required to have a mean diameter of $3\pm\SI{0.3}{\micro\metre}$ \cite{edana}, and specifies a cascade impactor be used to measure droplet size. Note that the standard does not specify what the distribution of droplet sizes is, but it does specify that the droplets are formed immediately before the mask, allowing little time for evaporation \cite{edana}. So we assume the droplets do not have time to dry out.
There is no test of fit to the face, so no requirement that a surgical mask fits well with few gaps for air to bypass the mask.

Another standard for masks is the F2100 standard of ASTM International \cite{F2100}. This has similar BFE requirements to the European Type II standard but in addition uses a test aerosol of $\SI{0.1}{\micro\metre}$ latex spheres. For example, the Level 3 standard F2100 standard requires that these particles must filter out with at least 98\% efficiency. This is in addition to a BFE of at least 98\%. So the requirements on the filtration properties are more stringent than for the Type II standard. However, there is still no test of fit in this standard \cite{F2100}.

Note that standards such as FFP2 (and N95) are designed to specify a minimum protection to all dangerous aerosols, not just droplets we breathe out that may contain a virus. So for example, they may also be worn when airborne asbestos is present. While the standards for surgical masks are aimed at reducing the amount of bacteria breathed out, for a wearer in an environment such as an operating theatre, that must be kept as sterile as possible.

\subsection{\joshspeaks{Focus of this work}}

While there is much literature on filtration, the details of how the size-dependent effectiveness of masks affect airborne transmission is less well understood.
There has been a vigorous debate on the role of particle size on transmission \cite{tellier2009,asadi2019,vuorinen2020,bourouiba2020}.
But this has focused on the size-dependent dynamics of particles in the air we breathe, rather than the competing effects that aerosol \joshspeaks{particle} size has on the viral dose versus the protection offered by a face covering.

Here we focus on the latter, and we aim to show the relative importance of mask fit and viral load on total protection.
We quantify the most important factors in mask protection by incorporating available data from the literature into a single-unified model describing:
\begin{enumerate}
\item How filtration depends on aerosol \joshspeaks{particle} size. Section \ref{sec:filtration} is largely review where we outline filtration for a general audience.
\item The size distribution of \emph{deposited} (fine) aerosol \joshspeaks{particles} $\lesssim$\SI{10}{\micro\metre} in diameter, encompassing both evaporation of \emph{exhaled} aerosols before inhalation and the probability that these are deposited in the respiratory tract upon inhalation.
\item How the viral load in the exhaler's respiratory fluid affects the size distribution of deposited viral aerosol \joshspeaks{particles}.
\item Finally, the expected overall reduction in deposited viral aerosol \joshspeaks{particles} from mask interventions under the combination of these effects.
\end{enumerate}
\joshspeaks{Note that we are interested in the overall protection offered to a community rather than just an individual; we therefore explore the effect of masking the exhaler (``source control'') in addition to the inhaler.}

We find that the final protection offered varies significantly depending on the mask material (including the number of layers in the cloth masks), the face seal and the degree of viral shedding.
The first factor has been reasonably well-explored during the pandemic \cite{zangmeister2020,konda2020,lustig2020,hao2020}, but the latter two factors have been underappreciated in our opinion.
A poorly fitted mask will offer limited protection, as aerosol \joshspeaks{particles} can bypass the mask material\cite{freitag2020,duncan2020}.
Finally, we show that the amount of viable virus prevented from being inhaled by a susceptible individual depends on the size distribution of respiratory \joshspeaks{particles} and the viral load of the exhaler.

\section{Masks are personal air filters}
\label{sec:filtration}

\begin{figure}
  \includegraphics[width=0.7\linewidth]{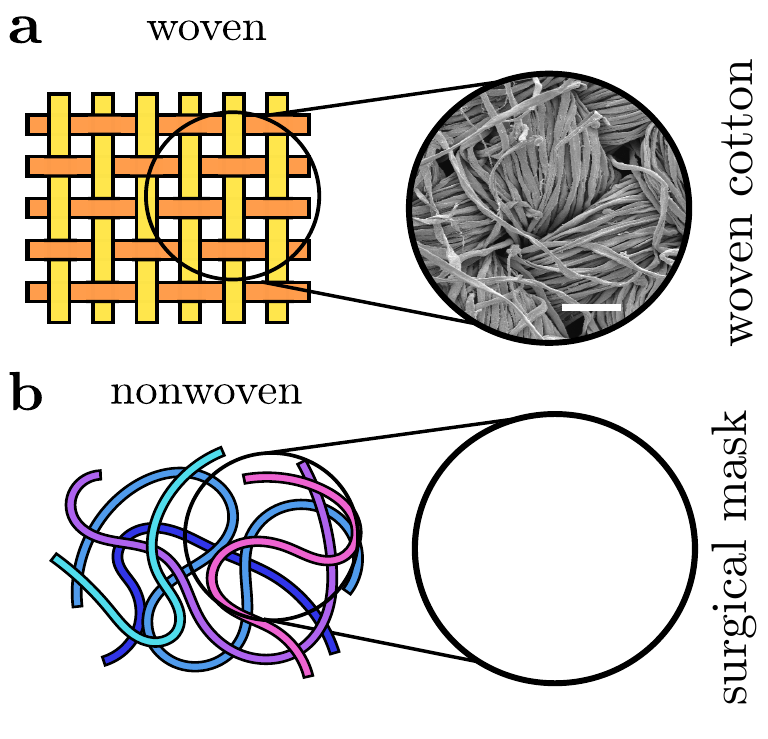}
  \caption{
    (colour online)
    Fabrics are broadly categorised as \emph{knitted} (not shown), \emph{woven} or \emph{non-woven}.
    (a) Woven fabrics formed by intersecting perpendicular yarns (the ``warp'' and ``weft'').
    (b) Nonwoven fabrics are formed by entangling fibres through other means, resulting in less ordered arrangements.
    Scanning electron microscope images of example fabrics show scalebars of (a) \SI{100}{\um} and (b) \SI{50}{\um}.
  }
  \label{fig:fabrics}
\end{figure}

Masks are air filters, and how these work and how efficient they are at filtering out particles of differing sizes is reasonably well understood \cite{robinson2021,wang2013}.
Coarser aerosol \joshspeaks{particles} with diameter $\gtrsim \SI{1}{\micro\metre}$, which are more capable of containing significant viral doses (\cf section \ref{sec:biological-factors}), are more easily filtered; finer \joshspeaks{particles $\sim$}\SIrange{0.1}{1}{\um} by contrast are transported around the fibre by the gas flow.
We illustrate this schematically in \Figref{fig:fabric-filter-efficiency}(a).

We show experimental measurements for the filtering efficiency in medical-grade respirators from the literature in \Figref{fig:fabric-filter-efficiency}(b).
Respirators are specialist materials typically composed of \emph{electret} fibres; these fibres \joshspeaks{carry} considerable electrostatic charge \cite{chen1993,kravtsov2000} which increases the efficiency in the \SIrange{0.1}{1}{\um} regime.
These charges diminish over time which will decrease the filtration efficiency.
The results for the base fabric (squares) are for its uncharged state, and thus suggest a lower bound on the respirator's efficiency after repeat use.

In \Figref{fig:fabric-filter-efficiency}(c) we have plotted both measurements of surgical mask filtering efficiency (symbols) and theoretical calculations (curves).
The theoretical calculations involved following the trajectories of particles inserted into the gas flow around fibres, described elsewhere \cite{robinson2021} and in the Supplementary Material (SM).
The efficiency is plotted as a function of the particle diameter, because the particle size ultimately determines how hard or easy it is to filter out.
Our model and the measurements of \Refscite{zangmeister2020} agree on the same basic facts:
\begin{itemize}
  \item Filtering efficiency is essentially \SI{100}{\percent} for particles $\gtrsim \SI{3}{\micro\metre}$ in diameter or larger.
  \item However, filtering efficiency is low (\SIrange{30}{60}{\percent}) in the range \SIrange{0.1}{1}{\micro\metre}.
\end{itemize}
These predictions make quantitative the picture we laid out in the preceding paragraph and \Figref{fig:fabric-filter-efficiency}(a).
Both surgical and cotton masks are thus only partially effective at filtering out sub-micrometre aerosol \joshspeaks{particles}.
However, their efficiency rapidly increases as the size increases beyond a micrometre, so masks are generally highly effective in this regime.
Note that there is considerable variation in mask quality, reflecting the many available standards.
The ``dental masks'' A and C of \Refcite{oberg2008} (shown in \Figref{fig:fabric-filter-efficiency} for reference) do not pass any mask standard (not even the less stringent BFE test at \SI{3}{\micro\metre}) making them of very poor quality.
Conversely, most of the surgical masks tested in \Refscite{oberg2008,duncan2020} outperform the predictions of our theoretical profile and \Refcite{zangmeister2020}.

For reference, SARS-CoV-2 is approximately $\SI{0.1}{\micro\metre}$ in diameter \cite{bar-on2020}, so any particle larger than this \emph{can} potentially carry a virus.
However, in the next section we will argue that only particles larger than $\gtrsim \SI{1}{\micro\metre}$ are \emph{likely} to contain any virus in the majority of cases; thus filtration efficiency in this regime is sufficient to significantly reduce transmission.

Surgical masks are meshes of fibres (\cf \Figref{fig:fabrics}(b)), whereas cloth face coverings are typically more ordered (\eg woven fabric in \Figref{fig:fabrics}(a)).
In \Figref{fig:fabric-filter-efficiency}(d) we compare measurements and model predictions for the material of two masks formed from 4 layers of plain-woven cotton fabrics \Refcite{zangmeister2020}.
There is considerable variation, but for the best fabric the behaviour is more-or-less identical to surgical masks; this is broadly in agreement with the findings of \Refscite{zangmeister2020,konda2020,lustig2020,wangData2020} and
demonstrates that reusable cloth masks can be suitable replacements for disposable surgical masks.

Returning to the mechanism, we briefly discuss the physics underlying this behaviour (more details can be found in \Refcite{robinson2021}).
Masks are fundamentally arrays of fibres, see \Figref{fig:fabrics}, so air must flow around and between these fibres.
Particles a fraction of a micrometre in size have very little inertia and so tend to follow the air flow through the mask faithfully avoiding the fibres.
However, the particle inertia varies with its mass \ie the cube of its diameter (its volume).
This means it rapidly increases with diameter.
Beyond around $\SI{1}{\micro\metre}$ in diameter the particles have too much inertia to follow the air as it twists in between the fibres, and so they impact onto the fibres.
On microscopic lengthscales most surfaces are attractive, so colliding particles will stick and remain on the fibres \cite{robinson2021}.
Because of this basic physics, the filtering efficiency of particles larger than $\sim\SI{3}{\micro\metre}$ is likely to be limited only by the leakage of air around the mask.
For the intermediate range \SIrange{1}{3}{\micro\metre} the exact behaviour will depend on the details of the material, but the rapid rise in filtration efficiency with particle size is a robust feature \cite{robinson2021}.
Finally, we note that the filtration efficiency increases for capture of the smallest aerosol \joshspeaks{particles} ($\lesssim \SI{0.3}{\um}$) in \Figref{fig:fabric-filter-efficiency}(b-c) where capture is enhanced by Brownian motion.
We have not focused on this mechanism because such small \joshspeaks{particles} are highly unlikely to carry significant doses of virus.

\begin{figure}
  \centering
  \includegraphics[width=\linewidth]{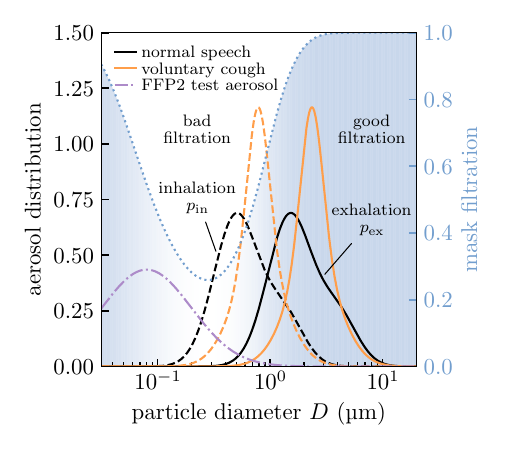}
  \caption{
    (colour online) Aerosol distributions relevant to masking: respiratory aerosols and the test aerosol used in FFP2 standards testing.
    We show the theoretical filtration profile of the surgical mask in \Figref{fig:fabric-filter-efficiency}(c) for reference.
    Aerosol \joshspeaks{particles} are coarser on exhalation (solid) than inhalation (dashed) because of evaporation, and so the exhaled size distribution overlaps more strongly with the region where masks filter effectively (dotted line and blue shaded region).
    The test aerosol (dash-dotted) contains primarily finer particles than are generally \joshspeaks{found} in respiratory aerosols.
  }
  \label{fig:size-dependence}
\end{figure}

\section{Size distribution of fine respiratory aerosol \joshspeaks{particles} involved in airborne transmission}
\label{sec:model}

To assess the effect of mask wearing we need to know which respiratory aerosol \joshspeaks{particles} are likely to carry significant doses of virus.
We thus need to know the size distribution of \joshspeaks{particles on exhalation and inhalation}, and how concentrated the (viable) virus is in aerosols produced by shedding in an infected individual.

\subsection{Size distribution of \joshspeaks{bio}aerosol \joshspeaks{particles}}

\joshspeaks{We need the size distribution of particles on inhalation \emph{and} exhalation. The former are obtained from the latter through knowledge of the evaporation kinetics, so we begin with exhaled particles.}

Expiratory particles can be meaningfully categorised by their site of origin in the respiratory system.
Larger \joshspeaks{particles} are more likely to be deposited in the respiratory tract \cite{zhang2012,cheng2014,chalvatzaki2020,guo2020}, so a large aerosol \joshspeaks{particle} originating deep in the respiratory tract would immediately deposit and be unlikely to escape; \joshspeaks{therefore} as a rule, smaller \joshspeaks{exhaled} aerosol \joshspeaks{particles} emerge from lower in the respiratory tract.
The majority of aerosol \joshspeaks{particles} (and droplets) produced in the oral cavity vary in size from $\sim$\SIrange{10}{1000}{\um} \cite{johnson2011}, whereas \joshspeaks{particles} produced in the Larynx and the lower respiratory tract are seen in the range $\sim$\SIrange{0.1}{10}{\um} \cite{morawska2009,johnson2011,asadi2019,gregson2021}.
The former presumably contain the majority of virus because volume scales as the diameter cubed \footnote{The concentration of viable virus may be lower in oral-mode droplets: in influenza it is a factor of $~10$ smaller \cite{milton2013}, however even accounting for this a factor of 10 increase in diameter increases the expected number of virions by $\sim 100$.}, but the latter will be our focus for reasons laid out in the introduction.

Fluid particles immediately begin to evaporate upon exhalation \cite{xie2007,liuCoughJets2017} and aerosol \joshspeaks{particles} smaller than $\lesssim \SI{10}{\um}$ will reach their dessicated steady states (which we will refer to as ``nuclei'') in less than \SI{1}{\second} \cite{nicas2005}.
We take parameterisations of the measured size distributions (\ie uncorrected for evaporation) reported in \Refscite{johnson2011,gregson2021} as the unperturbed (by masking) inhaled distribution $p_\mathrm{in}^{(0)}$ in our calculations.
Their hydrated state would be around a factor of $\sim3$ larger on exhalation, allowing for coverings to filter aerosol \joshspeaks{particles} more effectively at the source \cite{liuCoughJets2017,robinson2021}.
We therefore write the distribution of exhaled \joshspeaks{particle} diameters $D$ as
\begin{equation}
  p_\mathrm{ex}(D) \simeq p_\mathrm{in}^{(0)}(3 D).
\end{equation}
We show the \joshspeaks{size} distributions expected on inhalation and exhalation in \Figref{fig:size-dependence}.
We see that the bigger aerosol \joshspeaks{particles} on exhalation more strongly overlap with the region where surgical masks filter effectively (without mask leakage).
In the same figure we also show the size distribution of FFP2 test aerosols%
\footnote{The FFP2 standard\cite{EuropeMaskStandard} specifies the test aerosol has a (number) median diameter between \SIrange{0.06}{0.10}{\micro\metre} with a geometric standard deviation between \numrange{2.0}{3.0}.
We therefore model the FFP2 test aerosol size distribution as a log-normal with a median of \SI{0.08}{\micro\metre} and a geometric standard deviation of 2.5.},
which are much finer on average.
NB: we work in log-space for $D$ (in \si{\micro\metre}), so these distributions are probability densities where $p_{\{\mathrm{in},\mathrm{ex}\}} \, d\ln{(D/\si{\micro\metre})}$ are the infinitesimal probabilities.
$p_{\{\mathrm{in},\mathrm{ex}\}}(D)$ therefore have no units.

\begin{table}
  \begin{ruledtabular}
  \begin{tabular}{lcc}
    & \multicolumn{2}{c}{Fraction removed by mask} \\
    \cline{2-3}
    Aerosol particles & By number & By mass \\
    \hline
    Normal speech (inhalation) & \SI{48.3}{\percent} & \SI{87.5}{\percent} \\
    Normal speech (exhalation) & \SI{86.3}{\percent} & \SI{99.3}{\percent} \\
    Voluntary cough (inhalation) & \SI{56.2}{\percent} & \SI{82.5}{\percent} \\
    Voluntary cough (exhalation) & \SI{94.7}{\percent} & \SI{99.3}{\percent} \\
    FFP2 test aerosol & \SI{58.5}{\percent} & \SI{65.1}{\percent} \\
  \end{tabular}
  \end{ruledtabular}
  \caption{
    Expected efficacy of \emph{perfectly fitting} surgical masks at removing aerosol \joshspeaks{particles} with various size distributions.
    We assume the theoretical filtration profile for the surgical masks shown in \Figref{fig:fabric-filter-efficiency}(c).
    In practice, mask leakage would reduce the fraction removed.
    These numbers are expected to be similar for (perfectly fitting) multi-layered cloth masks (\cf \Figref{fig:fabric-filter-efficiency}(d)).
  }
  \label{table:fraction-droplets-removed}
\end{table}

\begin{figure}
  \centering
  \includegraphics[width=\linewidth]{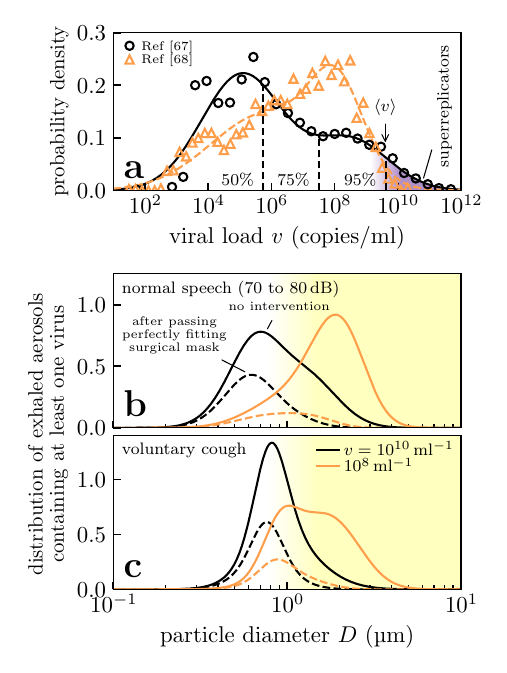}
  \caption{
    (colour online)
    Histograms of key properties of expiratory particles, \ie normalised so that the area under the curves gives the relative frequency.
    (a) Distributions of SARS-CoV-2 viral loads in testing from RT-PCR in two studies \cite{jones2020,jacot2020} (points) and our bimodal fits for calculating percentiles (lines).
    Note the large distribution, and the presence of a tail of patients with extraordinarily large viral loads (shaded purple) corresponding to so-called ``superreplicators''.
    (b-c) Aerosol distributions for virus-laden particles exhaled during speech and voluntary coughing under viral loads typical of the top 25th percentile in (a).
    We show the distributions of inhaled aerosols that contain at least one virus (solid lines) and those that bypass a \emph{perfectly fitting} surgical mask worn by the inhaler (dashed lines); the latter are unnormalised to show the effect of filtering.
    We calculate the former distributions using data from \Refscite{gregson2021,johnson2011} to characterise the exhaled aerosol \joshspeaks{particles} in healthy patients.
  }
  \label{fig:bioaerosol-distribution}
\end{figure}

\subsection{Perturbation of inhaled aerosol distribution by masks}
\label{sec:inhaled-distribution}

We write $\phi_{\{\mathrm{ex},\mathrm{in}\}}$ as the penetration through the mask (\ie the fraction of particles let through) on exhalation/inhalation.
We account for mask leakage $\alpha$ by setting
\begin{equation}\label{eq:mask-leakage}
  \phi(D; \alpha) = \alpha + (1-\alpha) \phi(D; \alpha=0)
\end{equation}
and we assume mask leakage to be independent of particle size $D$.
We obtain $\phi(D; \alpha=0)$ by subtracting the theoretical mask filtration shown in \Figref{fig:fabric-filter-efficiency}(c) from $1$.
Setting $\alpha = 1$ on inhalation or exhalation effectively removes the \joshspeaks{respective} mask within this model.
Mask leakage varies considerably, especially for surgical masks \cite{duncan2020}.
Estimates of $\alpha$ vary from around $\sim$\SIrange{20}{30}{\percent} in \Refscite{hossain2020,grinshpun2009} to as high as $\sim$\SIrange{40}{50}{\percent} in masks with \joshspeaks{artificial} leaks\cite{rengasamy2014}.
In general we expect $\alpha$ to be greater on exhalation than inhalation, because breathing affects the face seal, and therefore to vary with \eg breathing pattern \cite{grinshpun2009}.
For simplicity, we neglect these effects and keep $\alpha$ as a constant.
\joshspeaks{Initially we develop our model assuming $\alpha = 0$, before we consider the impact of leakage $\alpha > 0$ later in section \ref{sec:effectiveness}; for our current purposes it is enough to say that leakage effects are included in the definition of \eqref{eq:mask-leakage}.}

Accounting for masks, the inhaled size distribution becomes
\begin{subequations}\label{eq:total-mask-filtration}
  \begin{equation}
    p_\mathrm{in}(D)
    \simeq
    \frac{\phi_\mathrm{in}(D) \phi_\mathrm{ex}(3 D)}{1 - R_\mathrm{in}} p_\mathrm{in}^{(0)}(D),
  \end{equation}
  where the total (number) fraction of inhaled particles removed by the masks is
  \begin{equation}
    R_\mathrm{in} \simeq 1 -
    \int_0^\infty \phi_\mathrm{in}(D) \phi_\mathrm{ex}(3 D)
    \,
    p_\mathrm{in}^{(0)}(D)
    \, d\ln{\left(\frac{D}{\si{\micro\metre}}\right)}.
  \end{equation}
\end{subequations}
In Table \ref{table:fraction-droplets-removed} we show the results for the fraction of respiratory aerosol \joshspeaks{particles} removed by a perfectly fitting (\ie $\alpha = 0$) surgical mask assuming the theoretical filtration profile of \Figref{fig:fabric-filter-efficiency}(c).
We also show the results for an FFP2 test aerosol for reference, which involves finer \joshspeaks{particles} and so underpredicts the mask efficacy if one is really only interested in the respiratory aerosols.

The official FFP2 test measures effectiveness by mass (rather than number) weighting \cite{EuropeMaskStandard}.
Weighting $p_\mathrm{in}^{(0)}$ by $D^3$ in \eqref{eq:total-mask-filtration} gives the aerosol mass distribution, which may be more relevant for estimating inhaled viral dose, and the fraction of aerosol \joshspeaks{particles} removed by mass.
Mass weighting preferences larger \joshspeaks{particles} and improves the fraction removed in Table \ref{table:fraction-droplets-removed}, though this improvement is less pronounced for the FFP2 test aerosol because these coincide with the window where masks are poor filters.
An FFP2 test would fail this surgical mask which only filters \SI{65.1}{\percent} by mass, even though it would filter $\sim$\SI{99}{\percent} of the fine \joshspeaks{particles} exhaled during speech or coughing (with \emph{perfect} face seal).

\subsection{Accounting for biological factors: viral load and deposition}
\label{sec:biological-factors}

For an aerosol \joshspeaks{particle} to be a possible disease vector it must (i) contain at least one (viable) virion, and (ii) deposit in the respiratory tract upon inhalation.
We will model these two effects and show how they shift the relevant part of the inhaled size distribution to coarser aerosol \joshspeaks{particles}, which has positive implications for practical mask effectiveness.

Testing for SARS-CoV-2 is primarily performed by detecting the presence of viral RNA in respiratory fluid using reverse-transcriptase polymerase chain reaction (RT-PCR).
The distribution of concentrations of viral RNA, the \emph{viral load}, using this technique has been reported in \Refscite{jones2020,jacot2020} which we show in \Figref{fig:bioaerosol-distribution}(a).
The distribution is extremely broad, spanning around 10--12 orders of magnitude: patients in the upper tail of the distribution (the ``superreplicators'')  may be a factor in superspreading events.
Finally, we note that there is ample time for evaporation between sample collection and testing in these studies so we assume these concentrations to refer to the dessicated states (the nuclei).

Viral load typically peaks around the onset of symptoms \cite{zou2020} which is also the most contagious stage of disease progression \cite{anderson2020}; we thus expect viral loads from the upper half of the distributions in \Figref{fig:bioaerosol-distribution}(a) to be most relevant to disease transmission.
The actual number of \emph{viable} virus, as measured from viral plaque assays, in aerosol vs the RT-PCR result is typically only one part in \numrange{e2}{e4} \, \cite{milton2013,fears2020}.
We can thus take the upper limit of (naso-oral) viral loads as \SIrange{e8}{e10}{\per\milli\litre} instead of \SIrange{e10}{e12}{\per\milli\litre}.

These RT-PCR studies primarily involved testing naso- and oropharyngeal samples, which are presumably diluted by saliva and other fluids.
Fluids from the lower respiratory tract may be more concentrated with virus, especially when viral replication occurs there in advanced stages of the disease.
Viral loads larger than \SIrange{e8}{e10}{\per\milli\litre} may therefore be relevant to airborne transmission
We will find that factoring in viral load changes mask effectiveness in a sharp step function, and so any \emph{variation} in viral load through the respiratory tract is less important than where it is most concentrated.
As data on viral concentration in the lower respiratory tract is lacking, we can only discuss this possibility without factoring in the difference quantitatively.

\begin{figure}
  \centering
  \includegraphics[width=\linewidth]{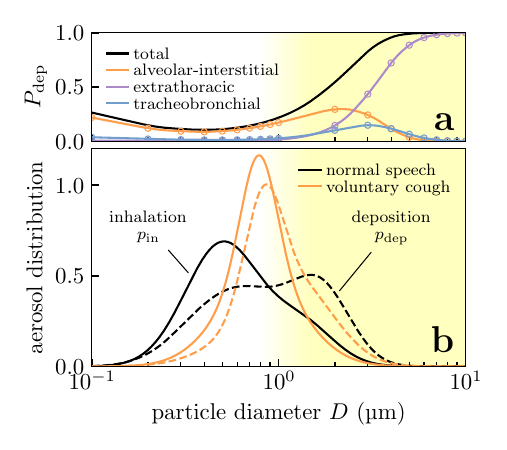}
  \caption{
    (colour online)
    \joshspeaks{Effect of deposition on the size distribution of aerosol particles received in the respiratory tract of the inhaler.}
    Fine aerosol \joshspeaks{particles} with diameters $\sim$\SIrange{0.1}{1}{\micro\metre} are unlikely to deposit in the respiratory tract, shifting the effective aerosol \joshspeaks{particle} size distribution of concern to coarser \joshspeaks{particles}.
    (a) Probability of deposition $P_\mathrm{dep}$ at various sites in the respiratory tract, showing experimental data (points) from \Refcite{chalvatzaki2020} and our quadratic interpolation (lines).
    (b) Probability density for inhaled bioaerosol \joshspeaks{particles} from \Refscite{gregson2021,johnson2011} (solid lines), and the probability density of those aerosol \joshspeaks{particles} which actually deposit in the respiratory tract (dashed lines) after conditioning on data in (a).
  }
  \label{fig:deposition}
\end{figure}

The probability that an aerosol \joshspeaks{particle} contains at least one virus depends on its volume and the viral load.
By combining this information we can estimate the fraction of particles at any specific size that contains the virus, and thus the variation in the relative number of particles that contain the virus across the size distribution.

The average number of virions $\langle n_v \rangle$ in a \joshspeaks{particle} is (assuming homogeneity) simply its volume times the viral load $v$ of the respiratory fluid.
This yields
\begin{equation*}
  \langle n_v \rangle = v \frac{\pi D^3}{6}.
\end{equation*}
assuming spherical \joshspeaks{particles}.
Assuming the Poisson distribution, the probability that a particle of this size contains at least one virion is
\begin{equation*}
  P_\mathrm{in}(v|D) = 1 - e^{-\langle n_v \rangle}.
\end{equation*}
This describes a step function in $D$ and $v$ with negligible probability for $D \lesssim \SI{1}{\um}$ and $v \lesssim \SI{e8}{\per\milli\litre}$.
$P_\mathrm{in}(v|D)$ is only $\sim$\SI{4}{\percent} for $D = \SI{1}{\um}$ and $v = \SI{e8}{\per\milli\litre}$, and rapidly diminishes further with decreasing $D$ or $v$.
Applying Bayes' theorem gives the size distribution of aerosol \joshspeaks{particles} containing at least one virus as
\begin{equation}\label{eq:inhale-distribution}
  p_\mathrm{in}(D|v) \propto P_\mathrm{in}(v|D) \, p_\mathrm{in}(D),
\end{equation}
with a proportionality constant to ensure the distribution is normalised.
The quantity $p_\mathrm{in}(D)$ is the distribution of \emph{all} inhaled aerosol \joshspeaks{particles} (\ie viral or otherwise) \joshspeaks{in the presence of masking \ie \eqref{eq:total-mask-filtration}}, that we considered in section \ref{sec:inhaled-distribution}.
In \Figref{fig:bioaerosol-distribution}(b-c) we show the resulting distributions produced in speech and coughing.
For the moderately large viral load of \SI{e8}{\per\milli\litre} the majority of viral aerosol \joshspeaks{particles} are distributed in the micron regime $\gtrsim \textrm{\SI{1}{\um}}$.
Only for extremely large viral loads of \SI{e10}{\per\milli\litre} do the submicron aerosol \joshspeaks{particles} begin to contain significant numbers of virus.

\joshspeaks{Our formalism focuses on the number distribution of aerosol particles, but it gives the mass distribution in the limit of small viral loads.
In this limit we find \[P_\mathrm{in}(v|D) = \langle n_v \rangle + \mathcal{O}(v^2),\] and so \eqref{eq:inhale-distribution} becomes
\begin{equation}\label{eq:inhale-distribution-mass}
  p_\mathrm{in}(D|v) \propto D^3 \, p_\mathrm{in}(D) + \mathcal{O}(v^2),
\end{equation}
\ie we obtain a mass-weighted inhaled distribution.
This is the appropriate weight function for calculating the inhaled mass (or dose), and so we can estimate the reduction in viral dose from masking as the small $v$ result in our model \eg in \eqref{eq:total-mask-filtration}.
}

As the second biological effect, we consider only those aerosol \joshspeaks{particles} which would actually deposit in the respiratory tract.
We show the probability of aerosol \joshspeaks{particles} being deposited in the respiratory tract $P_\mathrm{dep}$, using the data from \Refcite{chalvatzaki2020}, in \Figref{fig:deposition}(a).
We simply take $P_\mathrm{dep}$ as the total probability shown in \Figref{fig:deposition}(a), ignoring the site of deposition; we note however that the deposition site is potentially important in determining the severity of potential infection and so the model could be extended with further information on site-dependent risk.
Most of the fine aerosol \joshspeaks{particles} $\sim$\SIrange{0.1}{1}{\micro\metre} are simply re-exhaled after inhalation\cite{chalvatzaki2020}, which we can build into our model.
Applying Bayes' theorem again, we obtain the size distribution of inhaled aerosol \joshspeaks{particles} that deposit in the respiratory tract as
\begin{equation}
  p_\mathrm{dep}(D) \propto P_\mathrm{dep}(D) \, p_\mathrm{in}(D),
\end{equation}
with a proportionality constant to ensure the distribution is normalised.
This effect shifts the size distribution to coarser aerosol \joshspeaks{particles}, as shown in \Figref{fig:deposition}(b).

\emph{To summarise this section}, there is considerable biological variation in how the virus is shed.
Assuming the virus concentration depends only on the site of origin in the respiratory tract, then this concentration becomes independent of size for the fine aerosol \joshspeaks{particles}.
On physical grounds, we then expect the vast majority of the finest aerosol \joshspeaks{particles} $\lesssim \textrm{\SI{1}{\um}}$ to be empty of virus and/or to not actually deposit in the respiratory tract.
These fine \joshspeaks{particles} are therefore only expected to carry concerning doses in a small minority ($\lesssim \SI{5}{\percent}$) of infected individuals who shed orders-of-magnitude more virus than average.
Similar conclusions were recently reached by \citeauthor{freitag2020} \cite{freitag2020}.

\section{Overall mask effectiveness}
\label{sec:effectiveness}

\begin{figure}
  \centering
  \includegraphics[width=\linewidth]{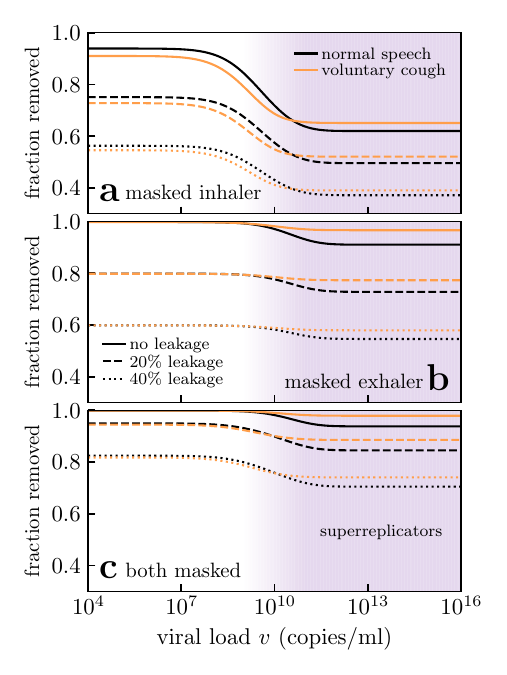}
  \caption{
    (colour online) Number fraction of viral aerosol \joshspeaks{particles} (\ie those containing at least one virus) from \joshspeaks{an} infected exhaler that are prevented from being deposited into the respiratory tract of the inhaler, considered where (a) only the inhaler is masked, (b) only the exhaler is masked and (c) both are masked.
    \joshspeaks{The upper limits at small viral loads correspond to the \emph{mass} fraction (\ie the depositing viral dose) removed.}
    In all three scenarios the effect of constant mask leakage is examined, and we assume: (i) a surgical mask with the theoretical filtration profile of \Figref{fig:fabric-filter-efficiency}(c), (ii) the bioaerosol \joshspeaks{particle} size distributions shown in \Figref{fig:bioaerosol-distribution}(b-c), and (iii) the deposition probabilities of \Figref{fig:deposition}(a).
    Biological variation affects the practical mask effectiveness, but the effect of mask leakage predominates.
  }
  \label{fig:total-effectiveness}
\end{figure}

To gauge overall mask effectiveness in preventing airborne transmission, we modify the (number) fraction of droplets removed \eqref{eq:total-mask-filtration} to focus on just those relevant to disease transmission.
That is, those aerosol \joshspeaks{particles} that (i) contain at least one virion and (ii) deposit in the respiratory tract of the inhaler. We presented those calculations in section \ref{sec:biological-factors} and show their effects on the \joshspeaks{particle} size distribution in \Figref{fig:bioaerosol-distribution} and \Figref{fig:deposition}.

Accounting for these effects, the aerosol \joshspeaks{particle} size distribution of concern becomes
\begin{subequations}\label{eq:total-mask-filtration2}
  \begin{equation}
    p_\mathrm{dep}(D|v)
    \simeq
    \frac{\phi_\mathrm{in}(D) \phi_\mathrm{ex}(3 D)}{1 - R_\mathrm{vec}}
    \frac{P_\mathrm{dep}(D) P_\mathrm{in}(v|D)}{1 - R_{\mathrm{dep},v}}
    p_\mathrm{in}^{(0)}(D),
  \end{equation}
  where the total (number) fraction of vector particles (\ie viral aerosol \joshspeaks{particles} that deposit) removed by the masks is
  \begin{equation}
  \begin{split}
    R_\mathrm{vec} \simeq 1 -
    \int_0^\infty \bigg( &
    \phi_\mathrm{in}(D) \phi_\mathrm{ex}(3 D) \, p_\mathrm{in}^{(0)}(D)
    \\
    & \frac{P_\mathrm{dep}(D) P_\mathrm{in}(v|D)}{1 - R_{\mathrm{dep},v}} \bigg)
    \, d\ln{\left(\frac{D}{\si{\micro\metre}}\right)},
    \end{split}
  \end{equation}
  and the (number) fraction of inhaled particles ignored because they do not deposit or contain virus is
  \begin{equation}
    R_{\mathrm{dep},v} = 1 -
    \int_0^\infty
    P_\mathrm{dep}(D) P_\mathrm{in}(v|D)
    \,
    p_\mathrm{in}^{(0)}(D)
    \, d\ln{\left(\frac{D}{\si{\micro\metre}}\right)}.
  \end{equation}
\end{subequations}
\Figref{fig:total-effectiveness} shows the net result of mask effectiveness $R_\mathrm{vec}$ against viral load under various masking scenarios.
In all scenarios the (number) fraction removed is seen to be a step function in decreasing mask effectiveness as (viable) viral loads vary over $\sim$\LineSIrange{e7}{e11}{\per\milli\litre}.
Masks are more effective with smaller viral loads because the virus-laden aerosol \joshspeaks{particles} are larger, as we saw in \Figref{fig:bioaerosol-distribution}(b-c).
In all cases this effect is further enhanced by the low-likelihood of deposition in the respiratory tract, accounting for which shifts the relevant size distribution to coarser aerosol \joshspeaks{particles} and enhances mask effectiveness.

\begin{figure}
  \centering
  \includegraphics[width=\linewidth]{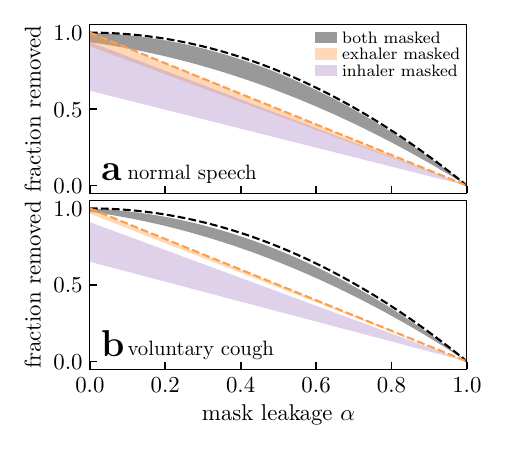}
  \caption{
    (colour online) Effect of mask leakage on (number) fraction of vector aerosol \joshspeaks{particles} (see text and \Figref{fig:total-effectiveness}) removed for two exhalation modes.
    The filled envelopes indicate the range of expected values from varying viral load as in \Figref{fig:total-effectiveness}.
    \joshspeaks{The upper limits are equivalent to the \emph{mass} fraction (\ie depositing viral dose) removed.}
    We show the expected performance for perfect filtration media (dashed lines) for two masks (black) and one mask (orange).
    We assume: (i) a surgical mask with the theoretical filtration profile of \Figref{fig:fabric-filter-efficiency}(c), (ii) the bioaerosol \joshspeaks{particle} size distributions shown in \Figref{fig:bioaerosol-distribution}(b-c), and (iii) the deposition probabilities of \Figref{fig:deposition}(a).
  }
  \label{fig:mask-leakage}
\end{figure}

As the effect of viral loads is seen as a step function, we plot the resulting range of values (\ie $R_\mathrm{vec}$ at small and large $v$) against mask leakage in \Figref{fig:mask-leakage}.
In all scenarios, masking the exhaler is better than masking the inhaler, because particles are larger on exhalation and  thus easier to filter; however, masking both is optimal.
Increasing mask leakage generally has a \joshspeaks{strongly} negative performance effect, especially on inhalation.
For mask leakage $\alpha$, the \emph{best} possible mask performance is $(1 - \alpha) R_\mathrm{vec}(\alpha=0)$ when one person is masked and $(1 - \alpha^2) R_\mathrm{vec}(\alpha=0)$ when both are masked; these theoretical values are shown as dashed lines in \Figref{fig:mask-leakage}.
However, \Figref{fig:total-effectiveness} and \ref{fig:mask-leakage} show that practical mask performance may be worse after considering the interplay of filtration with biological effects, especially when only the inhaler is masked.
This is because large viral loads increase the risk of transmission by submicron aerosol \joshspeaks{particles} where mask filtration is poor.

\joshspeaks{While we have focused on number fraction, the low viral load limit yields the expected result for the reduction in mass fraction (see discussion around \eqref{eq:inhale-distribution-mass}).
The mass fraction reduction estimates the reduction in viral dose transmitted.
Masks perform best in this limit, because coarser particles carry more mass.
The upper limit of the envelopes in \Figref{fig:mask-leakage} shows that in terms of viral dose reduction, masking almost performs as well as the ideal results (dashed curves) under conditions where the exhaler or both are masked.
If disease transmission depends on dose alone, then the filtration profile of the surgical mask in \Figref{fig:fabric-filter-efficiency}(c) is essentially perfect and the only limitations are (i) individual mask leakage and (ii) population adherence to mask wearing.
}

\section{Discussion}

We have seen that mask effectiveness crucially depends on the size of aerosol \joshspeaks{particles} which act as disease vectors, even for fine \joshspeaks{particles} $\lesssim \SI{10}{\micro\metre}$ in diameter.
At smaller viral loads, the virus-laden aerosol \joshspeaks{particles} are coarser making masks more effective; conversely, at large viral loads submicron \joshspeaks{particles} can carry virus which coincides with the size regime where mask materials have their poorest filtration properties.
The available data from \Refscite{jones2020,jacot2020} indicates that such large viral loads are extremely rare in the oral-naso fluid of SARS-CoV-2 patients, which would suggest that the coarser aerosol \joshspeaks{particles} $\gtrsim$\SI{1}{\micro\metre} are more important for disease vectors, and that masks would \emph{perform} better in practice than \eg an FFP2 test would predict.
However, we are unaware of any data on viral loads in fluid taken from deeper in the respiratory tract, where it may be more concentrated, and so submicron \joshspeaks{particles} remain potential transmission vectors.
Considering this key unknown, we are only able to set upper and lower bounds on the (number) fraction of viral aerosol \joshspeaks{particles} removed.

Taken together, our calculations suggest that mask effectiveness depends more on mask leakage than it does on the other variables.
Considering effectiveness for normal speech \Figref{fig:mask-leakage}(a), we find that the surgical mask of \citeauthor{zangmeister2020} with \SI{50}{\percent} \joshspeaks{mask leakage} would only filter \SIrange{62}{73}{\percent} of vectors (by number) when worn by both.
The upper limit (corresponding to low viral loads) is already close to the best possible performance of \SI{75}{\percent} obtainable through perfect filtration.
Improving the mask material is therefore insignificant compared to the benefit of improving the fit: keeping the material the same, but reducing leakage to \SI{5}{\percent} would yield mask performances of \SIrange{92}{99}{\percent}.
In a similar vein, ensuring both inhaler and exhaler are masked will outperform any gains in mask filtration when leakage is assumed; the lower limit of \SI{62}{\percent} at large viral loads already exceeds the theoretical best performance of a single mask of \SI{50}{\percent} at \SI{50}{\percent} leakage.

These findings have implications for surgical and fabric mask standards.
As discussed, the FFP2 test samples the submicrometer size regime which may not always be relevant for airborne transmission and thus may be too stringent as a test.
Conversely, the BFE test used for medical face masks featuring \SI{3}{\micro\metre} aerosol \joshspeaks{particles} may be too large and also sampling the region where mask material properties are likely to be poor; this may be too relaxed as a test.
Moreover, no standards for surgical and fabric masks include tests for fit, even though we have seen this to be a crucial limiting factor in their effectiveness at curbing disease transmission.

Finally, we note that our model is subject to the following limitations:
\begin{itemize}
  \item We only consider the number of aerosol \joshspeaks{particles}, \ie we are agnostic to the dose delivered or the dose-response.
  Our model is very conservative in this sense: weighting \joshspeaks{the} size-distributions by \joshspeaks{particle} volume to capture dose would increase the predicted effectiveness by selecting coarser \joshspeaks{particles} which masks are better at filtering.
  A consequence of this modelling choice is that mask effectiveness is a step function in the viral load.
  Focusing instead on dose would \emph{improve} the expected mask effectiveness\joshspeaks{, and in fact corresponds to the upper limit of effectiveness in our model (see discussion around \eqref{eq:inhale-distribution-mass}). Our stricter model allows for disease transmission to involve a kinetics depending on the number of virus-carrying particles received.}.
  Even with our conservative estimate, \Figref{fig:mask-leakage} shows that the filtration profile of surgical masks in \Refcite{zangmeister2020} is likely `good enough' to achieve almost ideal performance in practice is suggestive that (i) mask leakage and (ii) universal masking are more important factors than material properties.
  \item We assumed mask leakage to be independent of particle size.
  This would be straightforward to improve this by making $\alpha$ dependent on $D$ in \eqref{eq:mask-leakage}.
  \item We lack data on the viral loads in the lower respiratory tract.
  As a consequence we can only provide upper and lower bounds on mask effectiveness.
  \item Our mask effectiveness $R_\mathrm{vec}$ measures the reduction in exposure to potential disease vectors relative to when no masks are worn.
  As viral load increases, the absolute dose received (and thus the risk of transmission) increases even after $R_\mathrm{vec}$ has reached its plateau value (\cf \Figref{fig:total-effectiveness}).
  \item We have not focused on coarse aerosol \joshspeaks{particles} $\gtrsim$\SI{10}{\micro\metre} which have much shorter sedimentation times.
  Mask filtration is essentially perfect for these \joshspeaks{particles} so other factors related to masking are more important such as the way masks deflect and remove momentum from expiratory jets \cite{tang2009,tang2011}.
  \item We have assumed a constant flow velocity of \SI{6.3}{\centi\metre\per\second} through our masks to match \Refcite{zangmeister2020}.
  This neglects changes in flow velocity during tidal breathing, which will in turn be affected by the breathing pattern (and factors such as physical activity) which has a complex effect on mask performance \cite{grinshpun2009}.
  For simplicity we have neglected this, but an interesting extension of our model could consider the integrated effectiveness with variable flow velocity.
\end{itemize}

\section{Conclusion}

We have presented a model for practical mask performance, as measured by the (number) fraction of potential disease vectors removed.
Our model combines a mask's material filtration and leakage with three biological factors: (i) the distribution of respired aerosol \joshspeaks{particles}, (ii) the viral load of the exhaler and (iii) the probability of deposition in the respiratory tract of the inhaler.
We found that masks do protect the wearer, but perform best as source control; in any case, masking \emph{both} exhaler and inhaler is best.
But not all masks are the same. A mask meeting a Personal Protective Equipment (PPE) standard such as the European FFP2 standard should filter out at least 92\% of the virus.
The standard specifies 92\% filtration for a test aerosol that is smaller and harder to filter than the droplet sizes we expect to be most dangerous.

A surgical mask meeting the European Type II standard may be made of material which filters significantly less of the virus. For example, we predict that the material of one of the surgical masks studied by Zangmeister and coworkers \cite{zangmeister2020}, on inhalation filters 88\% of the mass of speech droplets, see Table \ref{table:fraction-droplets-removed}. The material of an FFP2 or similar mask will filter out close to 100\%. But the bigger problem with surgical mask standards is that they do not specify fit, and so many surgical masks fit poorly. Grinshpun and coworkers \cite{grinshpun2009} found that approximately 30\% of the air bypassed the material of the surgical mask they tested. With this poor a fit, the filtration is only $0.7\times 88\%=62\%$. This is enough to reduce transmission, but is inferior to an FFP2 mask.
Respirators offer both superior fit and superior filtration of smaller aerosol \joshspeaks{particles},
and so replacing surgical masks with FFP2, N95, or similar respirators will increase the protection of both healthcare workers and the public.

Following the emergence of more infectious variants of SARS-CoV-2, some policy makers have mandated the wearing of respirators in public spaces \cite{bavariaFFP2Mandate}.
As a complementary approach, policy makers could pursue a strategy of improving the quality of masks worn in community settings.
Practical guidance on reducing leakage would therefore be required to pursue this strategy.
For example, \citeauthor{duncan2020} \cite{duncan2020} found that surgical masks sealed via tie straps offered better face sealage than ear loops.
The filtration properties of fabric can be poor \cite{konda2020b,rios2021}, but their fit can be better than that of surgical masks \cite{duncan2020} and they can in principle be tailored to the wearer.
There are no standards for fabric masks apart from the recent F3502 standard \cite{F3502}, which does not set a standard for mask fit.
Washable cloth masks have the additional advantage of being more environmentally friendly than surgical masks and respirators, which are made from plastic fibres.

Surgical mask standards like the European Type II standard (EN 14683) are not adequate for a COVID-19. There is no requirement on mask fit, and filtration performance of the material is measured at the too-large diameter of $\SI{3}{\micro\metre}$. The standard could be made fit for purpose by specifying filtration as worn, as in the FFP2 standard, and measuring filtration at particle sizes around $\SI{1}{\micro\metre}$ or smaller. Alternatively, surgical mask standards could be removed entirely, leaving only FFP2 and similar standards. In either case, changing the standard could drive up the protection offerered by masks, and so reduce COVID-19 transmission.

Transmission of respiratory viruses is complex and poorly understood, so more data is needed.
We need either direct data on transmission rates as a function of conditions, with and without masks, and a much better idea of the infectivity of aerosolised virus including the required dose for infection.
Both of these will be challenging but both are possible.
The basic physics of filtration tells us about how capture varies with aerosol \joshspeaks{particle} size, and so once we have this data we can easily update our estimates of the protection offered by masks.

It now seems well established that with SARS-CoV-2 some infected people have viral loads thousands or millions of times higher than others \cite{jones2020,jacot2020}.
Thus a \SI{50}{\percent} reduction in dose due to mask wearing corresponds to very different absolute reductions in dose from infected people with high and low viral loads.
As typically the viral load of an infectious person will not be known, other forms of interventions are warranted in addition to masking.

\begin{acknowledgements}
  The authors wish to thank Kate Oliver for helpful discussions on textiles, Patrick Warren for guidance on LB simulations, Mahesh Bandi for making us aware of his ingenious use of a candyfloss maker, and Mike Allen, Jens Eggers, and Daan Frenkel for helpful discussions.
  We gratefully acknowledge Daniel Bonn, Patrick Charbonneau, K.\ K.\ Cheng, Rosie Dalzell, Tanniemola Liverpool, John Russo and Hajime Tanaka for providing valuable comments on this work. We would also like to thank the two reviewers for very useful and constructive comments on the submitted manuscript.

  JFR, JPR and CPR wish to thank the Bristol Aerosol COVID-19 group for valuable discussions and feedback on this work.
  JFR would like to thank Kirsty Wynne for assistance in debugging the code used in the theoretical calculations.
  The authors would like to thank Judith Mantell and Jean-Charles Eloi of the Wolfson Bioimaging Facility and the Chemical Imaging Facility (EPSRC Grant ``Atoms to Applications'', EP/K035746/1), respectively, for the SEM images and assistance in this work.

  We thank E. Chalvatzaki and M. Lazaridis for sharing their data on the probability of aerosol \joshspeaks{particle} deposition in the respiratory tract.
\end{acknowledgements}

\section*{Data availability statement}

\joshspeaks{The data presented as results in this study were obtained using a computer code which implements the methods outlined in the text. This} code \joshspeaks{is freely} available at \Refcite{maskflowGithub}.

\input{panic.bbl}

\printfigures
\printtables

\end{document}

%% file: panic.bbl
%